\documentclass[twocolumn,amsmath,amssymb,superscriptaddress,nofootinbib]{revtex4-1}
\pdfoutput=1

\usepackage{amssymb}
\usepackage{amsmath}
\usepackage{amsthm}
\usepackage[]{graphicx}
\usepackage[]{subfigure}
\usepackage{tensor}
\usepackage{color}
\usepackage{framed}
\usepackage{natbib}
\usepackage{dcolumn}
\usepackage{bm}

\usepackage{tikz}
\usetikzlibrary{decorations.pathmorphing, patterns,shapes}

\input epsf

\begin{document}

\allowdisplaybreaks
\begin{titlepage}

\title{
No smooth beginning for spacetime
}

\author{Job Feldbrugge}
\email{jfeldbrugge@perimeterinstitute.ca}
\affiliation{Perimeter Institute, 31 Caroline St N, Ontario, Canada}
\author{Jean-Luc Lehners}
\email{jlehners@aei.mpg.de}
\affiliation{Max--Planck--Institute for Gravitational Physics (Albert--Einstein--Institute), 14476 Potsdam, Germany}
\author{Neil Turok}
\email{nturok@perimeterinstitute.ca}
\affiliation{Perimeter Institute, 31 Caroline St N, Ontario, Canada}

\begin{abstract}
\noindent 
We identify a fundamental obstruction to any theory of the beginning of the universe, formulated as a semiclassical path integral. Hartle and Hawking's no boundary proposal and Vilenkin's tunneling proposal are examples of such theories. Each may be formulated as the quantum amplitude for obtaining a final 3-geometry by integrating over 4-geometries. We introduce a new mathematical tool - Picard-Lefschetz theory - for defining the semiclassical path integral for gravity. The Lorentzian path integral for quantum cosmology with a positive cosmological constant is meaningful in this approach, but the Euclidean version is not. Framed in this way, the resulting framework and predictions are unique. Unfortunately, the outcome is that primordial tensor (gravitational wave) fluctuations are unsuppressed. We prove a general theorem to this effect, in a wide class of theories. 
\end{abstract}
\maketitle
\end{titlepage}

In this Letter, we analyze two intriguing and longstanding proposals, due to Hartle and Hawking~\cite{Hartle:1983ai,Halliwell:1984eu} and Vilenkin~\cite{Vilenkin:1984wp,Vilenkin:1994rn}, respectively HH and V, describing the quantum creation of universes using the gravitational path integral. One is supposed to integrate over 4-geometries $g$ bounded by a final 3-geometry $h$. Formally, one writes
\begin{align}
{\rm HH}: \int^{h} [ dg] e^{-S_E[g] /\hbar} \qquad {\rm V}: \int^{h}_{\emptyset}[ dg] e^{i S[g] /\hbar}, 
\label{eq:LPLres}
\end{align}
where Hartle and Hawking advocate integrating over compact Euclidean 4-geometries bounded by $h$, whereas Vilenkin advocates integrating over Lorentzian 4-geometries interpolating between a vanishing initial 3-geometry, labelled $\emptyset$, and $h$. In the simplest case studied, the action is that for Einstein gravity with a positive cosmological constant $\Lambda$. Already in this case, there are interesting saddle point solutions. 

\begin{figure}[h] 
\begin{center}
\includegraphics[width=\linewidth]{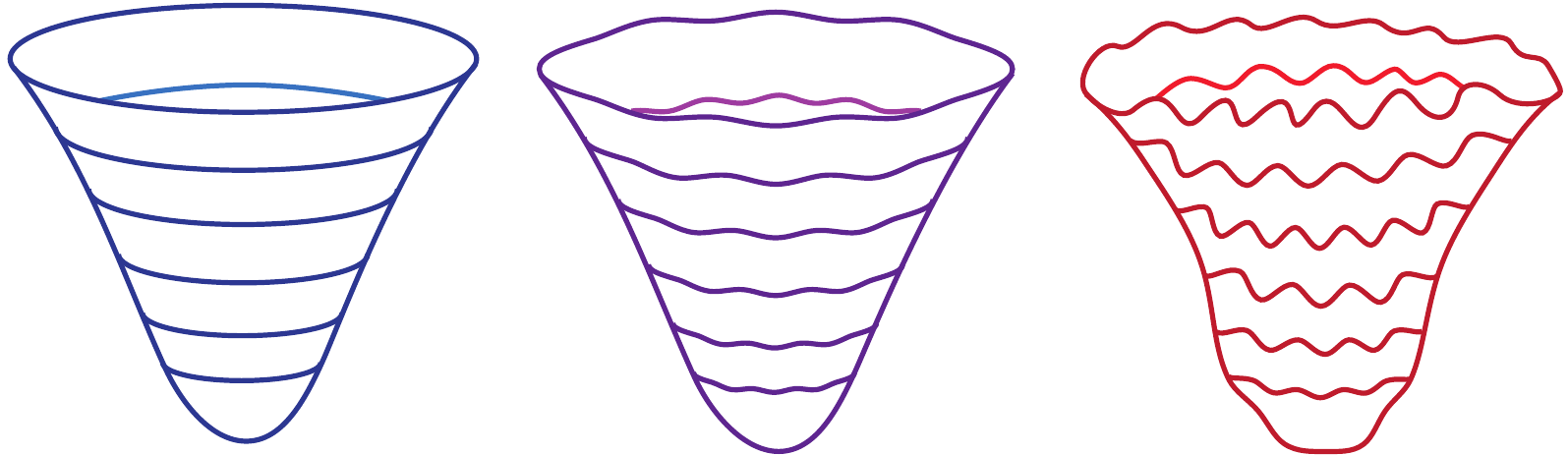}
\caption{\textit{Left:} the smooth, regular picture of the no boundary background. \textit{Middle:} the no boundary  picture with hoped-for small fluctuations, in agreement with observations. \textit{Right:} the fluctuations implied by the more rigorous, Lorentzian-Picard-Lefschetz approach developed here. Our analysis shows that, to leading semiclassical order, large fluctuations are preferred, leading to a breakdown of the theory.}
\label{fig:nbp}
\end{center}
\vskip -.7cm
\end{figure}

The new tool we bring to bear is Picard-Lefschetz theory, a powerful method for performing integrals like those in (\ref{eq:LPLres}) through steepest descent techniques~\cite{Arnold:singularities,Witten:2010cx,Feldbrugge:2017kzv}. We shall carefully analyze both the Hartle-Hawking and Vilenkin proposals by treating the 4-geometry as a homogeneous, isotropic cosmological background with gravitational waves described by general relativistic perturbation theory. The path integral is taken over all contributing 4-geometries, modulo diffeomorphism equivalence. In a suitable time-slicing, illustrated in Fig.~\ref{fig:nbp}, any topologically trivial 4-metric may be expressed as $-N(x^k)^2 dt^2 +h_{ij}(t,x^k) dx^i dx^i$, where $x^k$ are the space coordinates.  One may choose $t$ to run from $0$ to $1$, with the final 3-metric $h_{ij}(1,x^k)$. 

If the lapse $N$ is real, the four-geometry is Lorentzian; if imaginary, the four-geometry is Euclidean. Formally, one passes from the Lorentzian to the Euclidean theory with the replacement $N\rightarrow -i N\equiv N_E$, the sign being chosen to conform to the usual Wick rotation. Very generally, one cannot integrate $N_E$ over the infinite real range $-\infty<N_E<\infty$. Any real Euclidean action obtained from a real Lorentzian action is necessarily odd in $N_E$.  Furthermore, if its equations of motion are time-reversal invariant, they are even in $N_E$. Hence integrating out the dynamical variables always leaves one with an effective Euclidean action for $N_E$ which is odd in $N_E$. If it diverges to $+\infty$ as $N_E\rightarrow +\infty$, then it diverges to $-\infty$ as $N_E\rightarrow -\infty$, and vice versa. 
Therefore,  in any meaningful semiclassical Euclidean path integral, one cannot integrate $N_E$ over all real values. There are three available options: i) integrate $N_E$ over a half-range, should that integral converge; ii) leave the lapse real and Lorentzian, or iii) deform the lapse integral onto some other complex contour. We consider (and rule out) all three options. 

We perform the path integrals (\ref{eq:LPLres}) in the saddle point (semiclassical) approximation. First, we integrate over the background scale factor. Then, neglecting backreaction, we integrate over tensor (gravitational wave) perturbations. Both path integrals are Gaussian and present no difficulty. Finally, we carefully integrate over the lapse, using Picard-Lefschetz theory to identify the relevant saddle points and steepest descent contours~\cite{Feldbrugge:2017kzv}. 

Our key findings are as follows. First, for $\Lambda>0$ the Euclidean path integral diverges. Taken over $0^+<N_E<+\infty$, it diverges at $N_E=0$ due to the ``wrong sign" kinetic term for the scale factor. Integrating from $-\infty<N_E<0^-$ it diverges due to the cosmological constant. Thus we revert to the Lorentzian path integral which, taken either over real $0^+<N<\infty$, or  $-\infty<N<\infty$, yields a conditionally convergent, meaningful result. Picard-Lefschetz theory allows us to render the integral absolutely convergent by distorting the $N$ contour into the complex $N$-plane. We then obtain unambiguous predictions which are, unfortunately, unacceptable since they include unsuppressed perturbations on the final 3-geometry. Finally, we consider alternative complex contours for the lapse integral, including the recent proposal of Diaz Dorronsoro {\it et al.}~\cite{Dorronsoro:2017}. In a longer, companion paper \cite{Feldbrugge:2017b} we show that neither this, nor any other choice of contour avoids the problem of unsuppressed perturbations. We also study nonlinear back reaction numerically, showing it does not qualitatively alter our conclusions.  \\

In this Letter we shall neither explicitly discuss nor make use of the Wheeler-DeWitt equation. It seems to us that the path integral provides the most elegant, geometrical formulation of the Hartle-Hawking and Vilenkin proposals and therefore deserves investigation in its own right, through semiclassical methods. As already explained, one cannot define a semiclassical Euclidean path integral by integrating over $-\infty<N_E<\infty$. An important corollary is that no semiclassical Euclidean path integral can ever yield a solution of the homogeneous Wheeler-DeWitt equation (or a ``wavefunction of the universe"). See Ref.~ \cite{Feldbrugge:2017b} for further discussion. \\

There is a basic conundrum at the heart of quantum cosmology, whose resolution underlies our main claims. The problem is that the scale factor of the universe has a {\it negative} kinetic term, unlike all other degrees of freedom. This simple, but fundamental fact prevents one from Wick rotating time so that the phase factor $e^{i S/\hbar}$ appearing in Lorentzian path integrals becomes a real suppression factor $e^{-S_E/\hbar}$ for all degrees of freedom. Our approach is to perform no Wick rotation at all, but instead use Picard-Lefschetz theory to make sense of the original, Lorentzian path integral. In doing so, we uncover an important subtlety. For the simplest case of a closed, $\Lambda$ cosmology, the relevant saddle is a round Euclidean four-sphere, just as Hartle and Hawking claimed, but obtained via the conjugate continuation from de Sitter~\cite{Feldbrugge:2017kzv}. This inverts the semiclassical weighting, from $e^{+12\pi^2/(\hbar \Lambda)}$ to  $e^{-12\pi^2/(\hbar \Lambda)}$, agreeing with Vilenkin and representing a more physically intuitive $\hbar\rightarrow 0$ limit. 

However, the perturbations present new difficulties. The semiclassical amplitude is fixed by the complex, classical solution to the linearized Einstein equations giving the perturbation of the final 3-geometry. The Picard-Lefschetz construction ensures the convergence of the path integral and determines the prefactors uniquely. However, as a result of the abovementioned complex-conjugate nature of the background, the path integral yields an {\it inverse} Gaussian weighting for the final perturbation. Hence, large perturbations are favored and the theory is out of control. We give a general topological argument for this behavior, at the end of this Letter. 

To set the stage, we briefly review the path integral computation of perturbations in the flat slicing of a {\it classical} de Sitter background. The line element is $ a^2(\eta)(- d\eta^2 +d\vec{x}^2)$ with $a(\eta) = -1/(H \eta)$, (constant) Hubble parameter $H$ and conformal time $-\infty<\eta<0$. 
The Fourier modes of the perturbations decouple and can be treated independently. 
The quadratic action for a perturbation mode $\phi$ -- for example, a gravitational wave -- of wavenumber $k$ takes the form 
$S^{(2)}_{0,1}
=\frac{1}{2} \int_{\eta_0}^{\eta_1} \mathrm{d}\eta\, a^2(\eta) \left[ \left(\phi_{,\eta}\right)^2 -  k^2 \phi^2\right]$, with $\eta_0$ the initial and $\eta_1$ the final conformal time.
 We assume $|k\, \eta_0| \gg 1$ so that the perturbations start out in the local adiabatic vacuum at some early time $\eta_0$. 
 For simplicity, we take $\eta_1\rightarrow 0^-$, so the mode ends up frozen, with its physical wavelength far outside the Hubble radius. The amplitude for a final perturbation $\phi_1$ is then given by
\begin{equation}
G^{(2)}_\phi[\phi_1] = \int \mathcal{D}\phi\, e^{iS^{(2)}_{0,1}[\phi]/\hbar-\frac{1}{2}k a_0^2 \phi_{0}^2/\hbar}\,,\nonumber
\end{equation}
where the action $S^{(2)}_{0,1}$ incorporates the boundary conditions $\phi(\eta_{0,1}) =\phi_{0,1}$, and the functional measure includes an integral over $\phi_0$. The second factor represents the initial (assumed) adiabatic ground state wavefunction.


The functional integral is Gaussian so the saddle point approximation is exact. Stationarizing with respect to $\phi_0$ and using the Hamilton-Jacobi equation $\partial S^{(2)}_{0,1}/\partial \phi_0=-\pi_\phi(\eta_0)=-a^2 \phi_{,\eta}(\eta_0)$, we find the saddle point solution to be ``negative frequency" at early times. Solving the perturbation equation $\phi_{,\eta\eta} - (2/\eta) \phi_{,\eta} + k^2 \phi = 0$, with the given boundary conditions, the classical solution is  $\phi \approx \phi_1 e^{ik\eta}\left(1- ik\eta \right)$. Evaluating the semiclassical exponent and carefully taking the limit $\eta_1\rightarrow 0^-$, we find 
\begin{equation}
G^{(2)}_\phi[\phi_1] \propto e^{-\frac{k^3}{2H^2}\phi_1^2 + i \frac{k^2}{2H^2\eta_1}\phi_1^2}\,.\label{eqforg}
\end{equation}
The probability density is determined by the modulus squared of the amplitude. The divergent phase (which physically represents the final momentum of the mode) disappears and we recover the familiar result of a scale-invariant power spectrum for $\phi_1$.

The same result can be obtained by analytic continuation from the Euclidean theory. First, we Weyl transform the line element to flat space, and $\phi$ to $\chi =a\phi$. After an integration by parts, the Lorentzian action becomes 
$S^{(2)}_{0,1}
=\frac{1}{2} \int_{\eta_0}^{\eta_1} \mathrm{d}\eta\, \left[ \left(\chi_{,\eta}\right)^2 -  (k^2-2/\eta^2) \chi^2\right]$. Now we pass to Euclidean time $X\equiv i\eta$ and 
 $S_E\equiv -iS$, obtaining $S_E=\frac{1}{2} \int_{ X_0}^{ X_1} \mathrm{d}X\, \left[ \left(\chi' \right)^2 +  (k^2+2/X^2) \chi^2\right]$, with $'\equiv d/dX$,
 {\it i.e.}, a positive Euclidean action.  We compute $G_\chi[\chi[X_1]]$ from the Euclidean path integral over $\chi$. Again, we seek a classical saddle point solution. Finiteness of $S_E$ imposes regularity at $X\rightarrow -\infty$,  automatically selecting the ground state wavefunction. The desired classical solution is $\chi(X)=\chi_1 f(X)/f(X_1),$ with  $f(X) =e^{k X}(1/X-k)$. The on-shell action is $S_E(X_1)=\frac{1}{2} \chi \chi'(X_1) = \frac{1}{2}\chi_1^2 f'(X_1)/f(X_1)$. We continue back to Lorentzian time by setting $X_1=i\eta_1$. Taking the limit $\eta_1\rightarrow 0^-$  again yields (\ref{eqforg}), with an additional phase generated from the change of variables from $\phi$ to $\chi$.

Let us now turn to a consistent semiclassical path integral treatment of {\it both} the background and the perturbations, in order to understand why this fails to yield the above-mentioned standard results. We assume a homogeneous and isotropic background cosmology: $ds^2= - N_p(t_p)^2 \mathrm{d}t_p^2 + a(t_p)^2 \mathrm{d}\Omega_3^2, $ with lapse function $N_p$, scale factor $a(t_p)$ and unit $3$-sphere metric $\mathrm{d}\Omega_3^2$. The time $t_p$ is the physical time if $N_p$ is set to unity. The Einstein-$\Lambda$ action for the background is 
\begin{align}
S^{(0)}_{0,1} 
&= 2 \pi ^2 \int_0^1  \left[ - 3 a \frac{ a_{,t_p} ^2 }{N_p} + N_p( 3a - a ^3 \Lambda )\right]\mathrm{d}t_p \label{eq:Action}\,,\nonumber
\end{align} 
(in units where $8\pi G = 1$). The path integral to evolve from $a(0)=0$ to $a(1)=a_1$ is \cite{Teitelboim:1982,Halliwell:1988wc}
\begin{align}
G^{(0)}[a_1;0] = \int_{0^+}^\infty \mathrm{d}N \int_0^{a_1}\mathcal{D}a\, e^{i S^{(0)}[a,N]/\hbar}\,.\nonumber
\end{align} 
Re-defining the lapse and the time coordinate via $N_p \,\mathrm{d}t_p \equiv (N\,\mathrm{d}t)/a$ renders the action quadratic in $q\equiv a^2$,
\begin{equation}
S^{(0)}= 2 \pi ^2 \int_0^1 \left[ -\frac{3}{4 N}\dot{q}^2 + N(3 -  \Lambda q) \right]\mathrm{d}t \,. \label{ActionH}
\end{equation} 
The path integral over $q$ can now be performed exactly\footnote{Modulo issues regarding operator ordering and the path integral measure, and the restriction $q\geq 0$, further discussed in  \cite{Feldbrugge:2017kzv, Gielen2}.}. The classical solution satisfying $q(0)=0$, $q(1)=q_1$ is
\begin{equation}
q(t) =\frac{\Lambda}{3} N^2 t^2 + \left[q_1 - \frac{\Lambda}{3}N^2 \right] t\, .\nonumber
\label{qsol}
\end{equation}
The propagator reduces to:
\begin{eqnarray}
G^{(0)}[q_1;0] &=& \sqrt{\frac{3\pi i}{2\hbar}}\int_{0^+}^\infty \frac{\mathrm{d} N}{N^{1/2}} e^{i S^{(0)}[q_1;0;N]/\hbar} \label{eq:Nintegral}\,; \nonumber \cr
S^{(0)}[q_1;0,N]&=&2 \pi^2\left(N^3 \, \frac{\Lambda^2}{36} + N (3 -{1\over 2} \Lambda\, q_1) -\frac{3 q_1^2}{4N}\right) \,.\nonumber
\end{eqnarray}
This integral is then evaluated by deforming the integration contour into the complex $N$-plane, using Picard-Lefschetz theory~\cite{Arnold:singularities,Witten:2010cx} to identify the relevant saddle points and steepest descent contours. 

The on-shell background action $S^{(0)}[q_1;0,N]$ has four saddle points, each located in a different quadrant of the complex $N$-plane. The relevant saddle is located at
\begin{align}
N_s &= \frac{3}{\Lambda}\left(i + \sqrt{\frac{\Lambda}{3} q_1 - 1}\right),\nonumber
\end{align}
yielding for the no boundary propagator
\begin{equation}
G[q_1;0] \propto  e^{-\frac{12\pi^2}{\Lambda}   - i4\pi^2 \sqrt{\frac{\Lambda}{3}}\left(q_1-\frac{3}{\Lambda}\right)^{3/2} }\,.\nonumber
\end{equation}
As discussed in \cite{Feldbrugge:2017kzv,Feldbrugge:2017b}, Picard-Lefschetz theory implies semiclassical suppression, in agreement with Vilenkin but not with Hartle and Hawking.  

We have performed the analogous calculation with a slow-roll inflaton field $\varphi$ whose potential is well-approximated by $V(\varphi)\approx \Lambda - {1\over 2} m^2 \varphi^2$ near $\varphi=0$. We find that, as one would naively expect, for small $\varphi_1$,
\begin{equation}
G[q_1,\varphi_1;0,0] \propto  e^{-\frac{12\pi^2}{V(\varphi_1)} }\times phase \nonumber
\end{equation}
so there is a higher weighting for a larger initial potential energy $V(\varphi)$. Given that the radius of the universe is approximately $\sqrt{3/V(\varphi)}$ when space and time become classical, this supports the intuition that it is easier to nucleate a small rather than a large universe.

The same results can be obtained in physical time $t_p$ using the correspondence 
\begin{equation}
\sinh(H t_p) = H^2 \,N \,t - i\,, \label{eq:corres}
\end{equation}
where we define $H=\sqrt{\Lambda/3}$ and $a(t_p) = \frac{1}{H}\cosh\left(Ht_p\right)$. The no boundary point $t=0$ corresponds to $Ht_p = -\frac{\pi}{2}i.$ \\


Let us now extend our analysis to include perturbations -- for example, gravitational waves -- treated at leading (quadratic) order. The full propagator is
\begin{align}
G[q_1,\phi_1;0] = \int_{0^+}^\infty \mathrm{d}N \int ^{q_1}\mathcal{D}q\int^{\phi_1} \mathcal{D}\phi\, e^{i S/\hbar}\,,\nonumber
\end{align} 
where $S=S^{(0)}[q;0,N]+S^{(2)}[q,\phi,N]$, with 
\begin{equation}
S^{(2)}[q,\phi,N]=\frac{1}{2} \int N\mathrm{d}t\,\left[ q^2 \left(\frac{\dot{\phi}}{N}\right)^2 -  l(l+2) \phi^2\right],\nonumber
\end{equation}
and $l$ the principal quantum number on the 3-sphere. For notational economy we explicitly include just one, orthonormalized mode $\phi$; all modes occur in similar fashion.  For tensor perturbations, $l\geq 2$~\cite{Gerlach:1978gy}. (In general, one may also have scalar or vector perturbations, with $l\geq0$ and $l\geq 1$ respectively: see, {\it e.g.}, Ref.~\cite{Gratton:2001gw}).  The lapse perturbation is nondynamical in the absence of matter and may be set to zero. The no boundary condition is then implemented by specifying $q(0)=0$ and requiring the action to be finite and stationary under all variations which vanish on the final boundary. 

The path integral over the perturbations is again quadratic, so the saddle point approximation gives the $\phi_1$ dependence exactly. The equation of motion for $\phi$ is $
\ddot{\phi} + 2\frac{\dot{q}}{q} \dot{\phi} + \frac{N_s^2}{q^2}l(l+2) \phi = 0$, where we use the saddle point $N_s$ of the background, neglecting backreaction. The finite action solution is
$\phi(t) = \phi_1 F(t)/F(1) $, with 
\begin{align}
F(t) =& \left( 1 + \frac{i}{H^2 N_s t - i}\right)^{\frac{l}{2}} \left( 1 - \frac{i}{H^2 N_s t - i}\right)^{-\frac{l+2}{2}} \nonumber \\ & \times \left( 1 - \frac{i(l+1)}{H^2 N_s t - i}\right) \,. \label{eq:F}
\end{align}
Note $\phi(t) \propto t^{\frac{l}{2}}$ as $t\rightarrow 0$, implying $\phi$ is regular there. 

The classical action for the perturbations reduces to a surface term on the final boundary,
\begin{align*}
&S^{(2)}[q_1,\phi_1,N_s] 
= \frac{1}{2} \int_0^1 \mathrm{d}t \frac{\mathrm{d}}{\mathrm{d}t}\left[ \frac{q^2}{N_s} \phi \dot{\phi}\right]
 = \frac{q_1^2}{2N_s}\phi_1^2\frac{\dot{F}(1)}{F(1)} \nonumber \\ 
& = \frac{\phi_1^2}{2} \left[ -\frac{l(l+2)}{H} \sqrt{q_1} -i \frac{l(l+1)(l+2))}{H^2}+ {\cal O}\left(\frac{1}{\sqrt{q_1}}\right)\right]\,.\nonumber
\end{align*}
The full propagator for the perturbed background factorizes at this order $G[q_1,\phi_1;0]=G[q_1;0] G_\phi[\phi_1;0]$, with
\begin{equation}
G_\phi[\phi_1;0] \propto e^{\frac{l(l+1)(l+2)}{2\hbar H^2}\phi_1^2} \times phase \nonumber
\end{equation}
corresponding to an {\it inverse} Gaussian distribution. 

In order to compare our results with the Bunch-Davies vacuum, we convert \eqref{eq:F} to conformal time $\mathrm{d}\eta = \mathrm{d}t_p/a.$ The physical time and the conformal time are related by $\tan\left(\frac{\pi}{4} + \frac{\eta}{2}\right) = \tanh\left(\frac{H t_p}{2}\right)\,,$
where $-\infty < t_p < \infty$ corresponds to $-\pi < \eta < 0.$ Thus, as $\eta\rightarrow 0$,
\begin{equation}
\sinh(Ht_p) = 2 \frac{\tan(\frac{\pi}{4} + \frac{\eta}{2})}{1 - \tan^2(\frac{\pi}{4} + \frac{\eta}{2})} \rightarrow -\frac{1}{\eta} + \frac{\eta}{3} + \frac{\eta^3}{45} + \dots\nonumber
\end{equation}
which, using \eqref{eq:corres}, leads to the late time approximation
\begin{equation}
\phi = \phi_1 \left[ 1 + \frac{1}{2}l(l+2) \eta^2 - \frac{i}{3} l(l+1)(l+2) \eta^3 + \dots \right]\,. \nonumber
\end{equation}
This is the late time expansion of the ``positive frequency'' mode function, confirming that the no boundary condition selects the ``wrong'' mode function as compared to the adiabatic ground state. \\

Having demonstrated our claim that the perturbations are out of control in the no boundary description of quantum de Sitter spacetime, we would like to establish how general the result is. To begin with, we shall consider a fluid more general than a cosmological constant, but which is still ``adiabatic'', namely, the background pressure $P$ is a function of the energy density $\rho$ so that there is a unique cosmological history parameterized by the scale factor $a$.  Furthermore, we assume this classical evolution results in a smooth ``bounce'' of the scale factor such as occurs in the closed slicing of de Sitter spacetime. 

From our discussion above, it is clear that the on-shell classical action is all that is needed to determine the semiclassical exponent in the quantum propagator both for the background and for the perturbations. In the no boundary solutions, $q=a^2$ runs from $q_0=0$ to $q_1$, a positive value. Thus $q$ itself may be used as a time coordinate. The Friedmann constraint allows us to express the  background line element as
\begin{equation}
ds^2=-{\mathrm{d}q^2\over 4 q\,(\frac{1}{3}\rho(q) q-1)}+q \,\mathrm{d}\Omega_3^2,
\label{qmetric}
\end{equation}
where we allow the energy density $\rho(q)$ to vary with $q$. 

Cauchy's theorem enables us to deform the time (or $q$) contour upon which we evaluate the classical action as long as it does not cross any singularity. In particular, we can deform it to one in which $q$ is real everywhere. The line element (\ref{qmetric}) is Lorentzian for $q>3/\rho(q) $ but Euclidean for $0\leq q<3/\rho(q)$, and is easily checked to be regular at $q=0$. At $q=3/\rho(q)$, where $q=q_B$, the real, Lorentzian solution ``bounces,'' and $q$ therefore ceases to be a single-valued time coordinate. Our complex saddle point solution (\ref{qsol}) passes {\it below} this point in the complex $q$-plane: it is precisely this topological fact which results in the suppression of the semiclassical amplitude, required by Picard-Lefschetz theory~\cite{Feldbrugge:2017kzv}. Using the Friedmann constraint, the classical action (\ref{ActionH}) gives $i S^{(0)}=-6\pi^2 i \int  \mathrm{d} q \sqrt{ \rho \,q/3-1}$. Since we start in the Lorentzian region we take the branch cut to run leftwards from the point $q_B$, the classical ``bounce." Continuing the $q$ integral below the branch cut to $q=0$, we obtain for the real part of the semiclassical exponent $-6 \pi^2 \int_0^{q_B}\sqrt{1-\rho \,q/3}$. For a cosmological constant $\rho(q) = \Lambda$, we obtain $-12 \pi^2/\Lambda$. Continuing  above the branch cut yields $+12 \pi^2/\Lambda$, Hartle and Hawking's result, which is inconsistent with Picard-Lefschetz theory. 

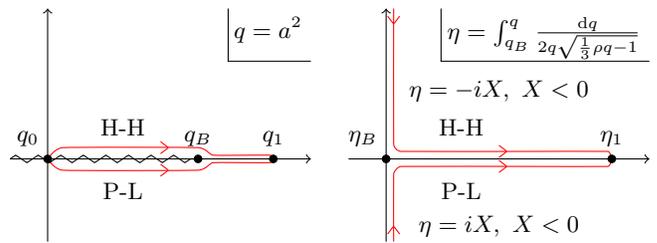
\begin{figure}[t] 
\centering
\begin{tikzpicture}

\draw [->] (-0.5,0) -- (3.5,0)  ;
\draw [->] (0,-1.1) -- (0,2)  ;

\draw[decoration = {zigzag,segment length = 3mm, amplitude = 0.5mm},decorate] (-0.5,0)--(2,0);

\draw [red] plot [smooth, tension=0.2] coordinates { (0,0) (0.2, 0.15) (2,0.15) (2.2,0.05) (2.95,0.05) (3.,0.0)};
\draw [red] plot [smooth, tension=0.2] coordinates { (1.5,0.23) (1.6,0.15) (1.5,0.07)};
\draw [red] plot [smooth, tension=0.2] coordinates { (0,0) (0.2, -0.15) (2,-0.15) (2.2,-0.05) (2.95,-0.05) (3.,0.0)};
\draw [red] plot [smooth, tension=0.2] coordinates { (1.5,-0.23) (1.6,-0.15) (1.5,-0.07)};

\draw (3.5,2) node  [below left] {$q=a^2$};
\draw [black] plot coordinates { (2.4,2) (2.4,1.3) (3.5,1.3) };

\draw (2,0.05) node [above] {$q_B$};
\draw[fill] (2,0) circle [radius=.05];

\draw (0,0.05) node [above left] {$q_0$};
\draw[fill] (0,0) circle [radius=.05];

\draw (3.,0.05) node [above] {$q_1$};
\draw[fill] (3.,0) circle [radius=.05];

\draw (1,0.2) node [above] {H-H};
\draw (1,-0.2) node [below] {P-L};


\draw [->] (4,0) -- (8,0)  ;
\draw [->] (4.5,-1.1) -- (4.5,2)  ;

\draw [red] plot [smooth, tension=0.1] coordinates {(4.6,2) (4.6,0.2) (4.7,0.1) (6.5,0.1) (7.45,0.1) (7.5, 0)};
\draw [red] plot [smooth, tension=0.2] coordinates { (6,0.18) (6.1,0.1) (6,0.02)};
\draw [red] plot [smooth, tension=0.2] coordinates { (4.52,1.9) (4.6,1.8) (4.68,1.9)};
\draw [red] plot [smooth, tension=0.1] coordinates {(4.6,-1.1) (4.6,-0.2) (4.7,-0.1) (6.5,-0.1) (7.45,-0.1) (7.5, 0)};
\draw [red] plot [smooth, tension=0.2] coordinates { (6,-0.18) (6.1,-0.1) (6,-0.02)};
\draw [red] plot [smooth, tension=0.2] coordinates { (4.52,-1) (4.6,-0.9) (4.68,-1)};

\draw (8,2) node  [below left] {$\eta = \int_{q_B}^{q} \frac{\mathrm{d}q}{2 q \sqrt{\frac{1}{3} \rho q -1}}$};
\draw [black] plot coordinates { (5.23,2) (5.23,1.3) (8,1.3) };

\draw (5.5,0.2) node [above] {H-H};
\draw (5.5,-0.2) node [below] {P-L};

\draw (6,0.9)  node {$\eta = -i X,\ X< 0$};
\draw (6,-0.9) node {$\eta = i X,\ X< 0$};

\draw (7.5,0.05) node [above] {$\eta_1$};
\draw[fill] (7.5,0) circle [radius=.05];

\draw (4.5,0.05) node [above left] {$\eta_B$};
\draw[fill] (4.5,0) circle [radius=.05];

\end{tikzpicture}
\caption{\textit{Left:} Analytic continuation contours (red) in the Hartle-Hawking (H-H) and Picard-Lefschetz (P-L) descriptions, above and below the branch cut in the complex $q$-plane. \textit{Right:} Corresponding contours for the conformal time.  }
\label{fig:Wick}
\vskip -.5cm
\end{figure}

To analyze the perturbations, we pass to coordinates in which the metric is conformally static: for $q>q_B$, we set $d\eta=dq/(2 q \sqrt{\rho \,q/3-1})$ to obtain the line element $q(\eta)(-d\eta^2+\mathrm{d}\Omega_3^2)$. We take $\eta=0$ to correspond to the ``bounce,'' so $\eta$ is positive in the Lorentzian region. Now, when $q$ passes below the branch cut commencing at $q_B$, the square root in the definition of $d\eta$ means that $\eta$ continues from the positive real $\eta$-axis onto the negative imaginary $\eta$-axis, $\eta=i X$ with $X<0$ in the Euclidean region.  Conversely, following $X$ forward from the Euclidean region, it  ``turns right'' into the Lorentzian region, whereas in the usual Wick rotation, assumed by Hartle and Hawking, it ``turns left'' (see Fig. \ref{fig:Wick}). Taking the continuation implied by Picard-Lefschetz theory for the background,  the Euclidean action for the perturbations has the ``wrong'' sign. We can still impose regularity of the modes in the Euclidean region, but the resulting semiclassical weighting factor will inherit the wrong sign.

As in our earlier discussion, it is convenient to go to a Weyl frame in which the kinetic terms are canonical. So we set $\phi=\chi/a$, obtaining for the Lorentzian action
\begin{equation}
i S^{(2)}=i \pi^2 \int d\eta \left[(\chi_{,\eta})^2+{a_{,\eta\eta}\over a}\chi^2 -l(l+2)\chi^2\right] \,.
\end{equation}
The background equations imply that $a_{,\eta\eta}/ a= {1\over 2} ({1\over 3} - w) \rho \,a^2 -1$, where $w=P/\rho$ is the equation of state. Analytically continuing $\eta$ back into the Euclidean region and then on to $q=0$ (corresponding to $X=-\infty$), as explained above, we obtain the Euclidean action
\begin{equation}
-S_E^{(2)}=\pi^2 \int_{-\infty}^0 d X \left[\chi'^2+U(X) \chi^2\right]\,,
\label{eucpertact}
\end{equation}
where $\chi'\equiv d\chi/dX$ and $U(X)\equiv l(l+2)+1 +{1\over 2} q\,(w_E-{1\over 3}) \rho_E $.  Here, $w_E$ and $\rho_E$ are the analytic continuations of their Lorentzian counterparts into the Euclidean region. Whatever the equation of state of the matter, $U(X)$ is positive at large $l,$ since regularity demands that $\rho_E$ remains finite, and correspondingly $w_E \rightarrow -1,$ as $q\rightarrow 0$. In fact, $U(X)$ is positive for all tensor modes as long as $\rho_E>0$ and $w_E>-17/3$. As before, the propagator's dependence on the final perturbation $\chi_1$ is given by the classical action. Finiteness of the action selects the mode $\chi=f(X)$ which is regular at $q=0$, {\it i.e.}, which vanishes at $X=-\infty$ (in the large $l$ limit, $f(X) \sim e^{\sqrt{l(l+2)}X}$). Using an integration by parts, from (\ref{eucpertact}) we obtain the on-shell Euclidean action $-S^{(2)}_E=\pi^2 \chi_1^2 f'(X_1)/f(X_1).$ The quantity $f'(X)/f(X)$ is positive at $X=-\infty$: as long as $U(X)$ is real and positive, the classical equation of motion for  $f$ implies $f'(X)/f(X)$  remains positive throughout the Euclidean region.

Continuing the conformal time into the Lorentzian region, we can show that the real part of the semiclassical exponent remains positive. Expressing the mode function in terms of its real and imaginary parts, $f(X)=R(X)+i I(X)$, we have shown that Re$[f'/f]=(R R'+II')/(R^2+I^2)>0$ at $X=0$. When $X$ turns in the negative imaginary direction, $X=-i \eta$, with $\eta$ positive, the Cauchy-Riemann equations yield $R'+iI'=i({R}_{,\eta}+i{I}_{,\eta}).$ Therefore, at $X=\eta=0$, we have ${R}_{,\eta}=I'$ and ${I}_{,\eta}=-R'$ and follows that the Wronskian $I{R}_{,\eta}-R {I}_{,\eta}$, which is independent of $\eta$, equals $(R^2+I^2)$Re$[f'/f]$ at $X=0$, which is positive. 
 Now, the real part of the semiclassical exponent, at a final Lorentzian time $\eta_1$ is similarly given, after an integration by parts, by $\pi^2 \chi_1^2$Re$[i\,{f}_{,\eta}(\eta_1)/f(\eta_1)]= \pi^2 \chi_1^2 (I{R}_{,\eta}-R {I}_{,\eta})/(R^2+I^2)$ (in fact, $I$ vanishes there by assumption). Since the Wronskian is positive, it follows that the semiclassical exponent for the perturbation $\chi_1$ is positive, for all positive $\eta$. 

In more general situations, the background pressure may not be expressible in terms of the density. In this case, it may not be possible to describe both the Euclidean and Lorentzian regions in terms of a real potential $U$. Nevertheless, even in this more general situation, where the ``bounce" point $q_B$ satisfying $q_B=3/\rho_B$ is complex, we still need to pass below it in the complex $q$-plane to be consistent with Picard-Lefschetz theory. This topological result again implies that the conformal time $\eta$ runs from $- i \infty$ in the region around $q=0$ to positive, nearly real values in an approximately ``Lorentzian" region. For modes of large $l$, the (in general complex) potential $U(X)$ is dominated by the $l^2$ term, and the no boundary solution is accurately described by the WKB Euclidean growing mode, so that Re$[f'/f]\sim \sqrt{l(l+2)}+O(l^{-1})$ at large $l$. The arguments above again demonstrate that the final semiclassical exponent has a positive real part. We conclude that the problem of unbounded perturbations, at small wavelengths, is unavoidable.  \\



\noindent {\it Acknowledgments:}
We thank Sebastian Bramberger, Claudio Bunster, Gary Gibbons, Steve Giddings, Steffen Gielen, Jonathan Halliwell, James Hartle, Thomas Hertog, Nick Manton and Alex Vilenkin for stimulating discussions and correspondence. Research at Perimeter Institute is supported by the Government of Canada through Industry Canada and by the Province of Ontario through the Ministry of Research and Innovation.

\bibliography{PicardLefschetzPerturbations}

\end{document}